\documentclass{aa} 
\usepackage{graphics} 
\usepackage{epsfig} 
\usepackage{amssymb}
\usepackage{txfonts}

\def\etal{et al. }  
  
\def\msun{{${\rm M}_\odot$} }

\def\Z0{{$\rm\,Z_\odot$}}

\def\n#1{{~}}
\def\spose#1{\hbox to 0pt{#1\hss}} 
\def\lta{\mathrel{\spose{\lower 3pt\hbox{$\mathchar"218$}} 
     \raise 2.0pt\hbox{$\mathchar"13C$}}} 
\def\gta{\mathrel{\spose{\lower 3pt\hbox{$\mathchar"218$}} 
     \raise 2.0pt\hbox{$\mathchar"13E$}}} 
\def\h0{} \def\q0{{q$\_0$}}

\def\Ha{${\rm H_\alpha~}$}

\def\OII{${\rm [OII]~}$}

\begin{document}

\title{Metallicity dependent calibrations of flux based SFR tracers} 
 
\author{Jens Bicker, Uta Fritze -- v. Alvensleben} 
 
\institute{Institut f\"ur Astrophysik, Universit\"at G\"ottingen, Friedrich-Hund-Platz 1, D--37077 G\"ottingen, Germany}

\offprints{ufritze@astro.physik.uni-goettingen.de} 
 
\date{Received ... , 2005; accepted ... , 2005} 
 
\authorrunning{J. Bicker \& U. Fritze -- v. Alvensleben} 
\titlerunning{} 
 
\abstract{  

We present new calibrations of the widely  used \Ha, \OII, and UV luminosity vs. star
formation rate (SFR) relations.  Using our evolutionary synthesis code GALEV we
compute the different calibrations for 5 metallicities, from 1/50 solar up to
2.5 solar. We find significant changes in the calibrations for lower
metallicities compared to the standard calibrations using solar input physics.    


}

%______________________________________________________________  

\maketitle 

\section{Introduction} 

Determining the Star Formation Rate (SFR) of galaxies is one of the most important steps to
understand their nature and evolution. This is usually done by
using the luminosity of the ${\rm H_\alpha}$ or the \OII line, as well as the
luminosity at 1500\AA~ and 2800\AA~ in the UV, as tracer of the ongoing star formation (SF)
(Kennicutt 1998, Gallagher \etal 1989, Madau \etal 1998, hereafter K98, G+89, M+98). These methods use
the fact that the UV continuum and the output of ionising Lyman continuum
 (Lyc) photons, responsible for the gaseous emission, are dominated by young massive stars
with masses ${\rm \gta 10~M_\odot} $ and lifetimes ${\rm \lta 2 \times 10^7~yr}$. 
 
Conventionally, the standard calibrations are derived on local samples of normal, i.e. big galaxies, assuming solar metallicity, as
appropriate. In the local universe, most of the actively star-forming galaxies are moderate to low luminosity late-type and dwarf galaxies having subsolar metallicities, and a significant amount is contributed by these systems to the local SFR density (e.g. Brinchmann \& Ellis 2000, Brinchmann {\sl et al.} 2004). Going to higher redshift, spectroscopic studies are 
clearly biased towards the brightest and most metal rich systems at every redshift. Even those, however, reveal substantially subsolar metallicities (e.g. Mehlert et al. 2003) 
and multi-band photometry of deep fields reaching the bulk of the intrinsically fainter galaxy population will clearly be dominated by subsolar metallicity galaxies and protogalaxies.

While the effects of the initial mass
function (IMF) (K98) or the absorption by dust are often discussed as sources of uncertainty in determining the SFR (e.g. Inoue \etal
2001), metallicity effects are rarely addressed. Concentrating on the SFR of star-bursting
dwarf galaxies derived from \Ha line fluxes, the metallicity dependence was examined e.g. by
Weilbacher \& Fritze - v. Alvensleben (2001). For a first extensive observationally based investigation into 
the metallicity dependence of \OII as a SFR indicator see Kewley \etal (2004).

We here investigate the effect of metallicity on the calibration of the various SFR
estimators, using our evolutionary synthesis code GALEV, described in Bicker
\etal (2004), extended to include the gaseous emission, for 5 metallicities from
1/50 solar up to 2.5 solar.

We first present some details of the gaseous emission as included into our code in Sect.
\ref{model_description}.  We then discuss the derived calibrations aacounting for the effects of metallicity as well as their dependence on the mass limits of the IMF and on the stellar evolutionary input physics in Sect. \ref{results}.  Finally we summarize our
results in Sect. \ref{conclude}.

\section{GALEV: galaxy evolution models}  
\label{model_description} 

Our chemically consistent evolutionary synthesis code GALEV is based on a modified version of  Tinsley's
equations for the chemical enrichment of the inter stellar medium and on isochrones from the Padova group (Bertelli \etal 1994) in the Nov.
1999 version, and the spectral library from Lejeune \etal (1997, 1998) for the spectral and photometric evolution of the stellar component.  For a
detailed description see Bicker \etal (2004). Now we have included the effects
of gaseous emission, in terms of lines as well as continuous emission. 
This was already included into our GALEV models for single bursts single metallicity stellar populations (SSPs) by Anders \& Fritze - v. Alvensleben  (2003), so we follow the method used
there to implement the gaseous emission into our chemically consistent galaxy models. To clarify the effects of metallicity we use galaxy models with fixed metallicities in this Letter.       
 
\subsection{Gaseous emission} 

The gaseous emission is dominated by very hot stars, which produce 
hydrogen ionising (Lyc) photons. This means, that the emission is
important in the early phases of an SSP (cf. Anders \& Fritze - v. Alvensleben  2003), or, in case of a galaxy, in stages of
(high) star formation activity.  For every star on every isochrone we calculate
the flux of Lyc photons from up-to-date non-LTE expanding model
atmospheres (Schaerer \& de Koter 1997, Vacca et al. 1996, Smith et al. 2002).
Summing up all Lyc photons of all stars on all isochrones at every timestep
gives us the total number of Lyc photons
(${\rm N_{Lyc}}$) at each time. On the basis of  ${\rm N_{Lyc}}$ we calculate
the gaseous continuum emission and the hydrogen line fluxes as described in
Kr\"uger \etal (1995) and Weilbacher \etal (2000). So the ${\rm H_\beta}$ flux is
given by:

\begin{displaymath}      
	{\rm F(H_\beta)=4.757 \times 10^{-13} \cdot N_{Lyc} \cdot f}, 
\end{displaymath}
where f describes the fraction of Lyc photons actually involved in
ionising the ISM.  For higher metallicities (${\rm Z \ge 0.008}$) we assume 
that 30\% (${\rm f=0.7}$) of the  Lyc photons are absorbed by dust immediately and cannot ionise the gas, i.e. ${\rm f=0.7}$ 
(Mezger 1978, Weilbacher \etal 2000). For the lower metallicities we take
${\rm f=1}$. Note that in this respect our models are not completely self-consistent. The input we use implies that the most massive stars at low metallicity rapidly emerge from their dust cocoons. Lines other than ${\rm H_\beta}$ are calculated from line ratios relative to ${\rm H_\beta}$.
Line ratios for hydrogen lines are taken from Stasinska's (1984) theoretical models, while for lines
of other elements we prefer to use empirical line ratios as derived from observations by Izotov \etal (1994, 1997) and Izotov \& Thuan (1998).
We include a total
of 31 hydrogen lines up to the Brackett series and 36 lines of
other elements. For the continuous emission and a detailed description see  Anders \& Fritze - v. Alvensleben (2003).
We recall that the metallicity dependence of the isochrones induces a
metallicity dependence for the flux of ionising Lyc photons that, in turn,
leads to a metallicity dependence of the hydrogen lines. For lines of other
elements, additional metallicity dependences come into play in a variety of
ways. Line ratios themselves depend on the metallicity of the gas in a specific
and complicated way for each element/line and differences in the typical
electron temperatures and densities in different metallicity environments add
to this. That is the reason why we use for these lines ratios derived from
observations of a large number of different galaxies at each of our subsolar
metallicity intervals.

In addition to young massive stars white dwarfs also have a significant output  of Lyc photons. 
Their emission, however, is associated with the planetary nebula phase, which is fairly short in comparison with our
minimum timestep ($\sim 25 \times 10^3$ yrs vs. $4 \times 10^6$ yrs). Hence, we decide to ignore
the contribution of the WDs to the total Lyc flux of actively star-forming galaxies. Binary stars are not yet included in our models at that stage.

\subsection{ZAMS extension of the Isochrones}
The youngest age provided by the Padova isochrones that we use is 4 Myr. This corresponds to an upper
mass limit of ${\rm \sim 70~M_\odot}$. For the gaseous
emission, however, the youngest and most massive stars play a very important role,
because the number of Lyc photons (${\rm N_{Lyc}}$) increases
dramatically with stellar temperature, ${\rm N_{Lyc} \sim  T_{eff}^4}$. The original isochrones hence unavoidably implied a severe underestimate of the Lyc flux. 

Stellar evolutionary tracks from the Padova group, however, start from zero age main sequence (ZAMS) and are available for stars up to 120 ${\rm
M_\odot}$. Hence we decided to supplement our set of isochrones by adding ZAMS isochrone. Stars on the ZAMS are not evolved, so there is no need to consider equivalent stellar evolutionary stages, usually a challenge in converting tracks to isochrones.
Now we are able to go to higher upper masses than the original isochrones
provided. 

For our models we now adopt an upper mass limit of 100 ${\rm M_\odot}$
as our standard and investigate 120 ${\rm M_\odot}$ for comparison. 

\section{Results}
\label{results}
To study the metallicity dependence of the various SFR tracers we calculate models with constant SFR
at the 5 available metallicities. We assume a Salpeter IMF with lower and upper mass limits of 0.15 and $100~{\rm M_\odot}$ as our standard and explore 0.1 and 
$120~{\rm M_\odot}$ for comparison. We have also explored different star formation histories (SFH) and found no differences in the calibrations as long as the SFR is evolving smoothly. 
For a discussion of rapidly changing SFRs, like in the short starbursts in dwarf galaxies, see Weilbacher \& Fritze - v. Alvensleben (2001). 

The derived calibrations for SFRs in terms of emission lines directly depend on
the fraction f of actually ionising Lyc photons. For comparison purposes, we
also give the values for ${\rm f=1}$ in the case of higher metallicities.  The
calibrations are not corrected to any amount of dust. Such a  correction
can easily be applied by calculating the ratio between the flux emerging from a
dusty galaxy ${\rm F_{dust}}$ and the corresponding unextincted flux ${\rm F_0}$. For any extinction
law ${\rm k(\lambda)}$ this is given by

\begin{displaymath}
	{\rm \frac{F_{dust}(\lambda)}{F_0(\lambda)} = 10^{-0.4 \cdot E(B-V)} \cdot k(\lambda)}.   
\end{displaymath}
Because of the wide range of extinction values and the variety of extinction laws appropriate for various types of galaxies (e.g. Calzetti 1997)
we prefer to give the calibrations uncorrected for dust to be applied to extinction corrected observed galaxy spectra. 

\begin{table}
\begin{center}
        \caption{Calibration constants for the  \Ha and \OII luminosity vs. SFR relations. (Values in brackets are for ${\rm f=1}$)}
%        \begin{footnotesize}
%	\begin{sf}
        \begin{tabular}{r|ll|ll}\hline\hline
	 \multicolumn{1}{c}{${\rm M_{up}}$} & \multicolumn{2}{|c|}{100~\msun} & \multicolumn{2}{c}{120~\msun}\\
        \hline
	 Z& $\rm C_{H_\alpha}$& $\rm C_{[OII]}$ & ${\rm C_{H_\alpha}}$& ${\rm C_{[OII]}}$\\
	\multicolumn{1}{c}{} & \multicolumn{2}{|c}{$\left[{\rm \times 10^{41}\frac{erg ~ s^{-1}}{M_\odot ~ yr^{-1}}}\right]$} & \multicolumn{2}{|c}{$\left[{\rm \times 10^{41}\frac{erg ~ s^{-1}}{M_\odot ~ yr^{-1}}}\right]$}\\
	
        \hline
          0.0004 & 2.7       & 0.5       & 3.0       & 0.5      \\
          0.004  & 2.3       & 1.4       & 2.6       & 1.6      \\
          0.008  & 1.4 (2.1) & 1.5 (2.2) & 1.7 (2.3) & 1.8 (2.6)\\
          0.02   & 1.3 (1.8) & 1.3 (1.9) & 1.6 (2.2) & 1.6 (2.3)\\
          0.05   & 1.2 (1.7) & 1.2 (1.8) & 1.4 (1.9) & 1.4 (2.0)\\
        \hline
       \end{tabular}
%       \end{sf}
%	\end{footnotesize}
       \label{calib1}
\end{center}
\end{table}

\subsection{\Ha}

The standard calibration for SFRs derived from \Ha luminosity is given by K98: 

\begin{displaymath}
{\rm L_{H_\alpha}~(erg~s^{-1}) = C_{H_\alpha} \cdot SFR_{H_\alpha}~(M_\odot ~ yr^{-1})} 
\end{displaymath}
K98 derives ${\rm C_{H_\alpha}= 1.3 \times 10^{41}}$ using population synthesis
models with solar abundances and a Salpeter IMF with an upper mass limit of 100~${\rm
M_\odot}$. For our solar metallicity model we find exactly the same value for  ${\rm C_{H_\alpha}}$, but we also find a strong dependence of ${\rm C_{H_\alpha}}$ on
metallicity of more than an factor of two between our solar metallicity and our lowest metallicity model (1/50
solar) (cf. Tab. \ref{calib1}), even though the \Ha line strength itself does not directly depend on the gas
metallicity. The effect is caused by the higher temperatures and, hence, higher ionising fluxes of low metallicity stars as compared to higher metallicity stars. Using the standard calibration for solar abundance will overestimate the
SFR in low metallicity galaxies by up to a factor $\geq 2$. 

If we choose 120~${\rm M_\odot}$ as our upper
mass limit, model galaxies get higher Lyman
continuum photon fluxes, resulting in stronger emission lines. Therefore, the
calibration constant increases by 10 $-$ 15\%, slightly depending on metallicity, as seen in Tab. 1., resulting in lower
SFRs at a given \Ha luminosity.

\subsection{\OII}
The forbidden \OII line at 3727 \AA~ is often used  as a SF indicator for galaxies at higher
redshift, where the \Ha line is redshifted out of the optical window.
Calibrations by G+89 and Kennicutt (1992) yield

\begin{displaymath}
{\rm L_{[OII]}~(erg~s^{-1}) = C_{[OII]} \cdot SFR_{[OII]}~(M_\odot ~ yr^{-1})},
\end{displaymath}
with ${\rm C_{[OII]}= 1.5 \times 10^{41}}$ and $5 \times 10^{40}$, respectively. Both calibrations are based on observed galaxy samples with SFRs previously derived from \Ha. Differences in the galaxy samples are responsible for the differences in the calibration constants. G+89 used blue irregular galaxies, 
Kennicutt (1992) a sample of normal, mostly spiral-type galaxies. They 
probably reflect the lower extinctions and metallicities in irregular
galaxies as compared to normal spirals. 
K98 derived a corrected (IMF, \Ha calibration) average for the calibration constant of ${\rm C_{[OII]}= 0.7 \times 10^{41}}$. 
Our model results are close to those of
G+89, which better suit to our dust-free models (cf. Tab.~\ref{calib1}).  With increasing 
metallicity we find a rise in the calibration factor up to half-solar
metallicity. The first step in metallicity from Z=0.0004 to 0.004 alone 
increases ${\rm C_{[OII]}}$ by a factor 3.
This reflects the increasing  oxygen abundance.  Towards solar and
super-solar abundances, however, the calibration factor drops as a result
of the lower output of ionising photons of high metallicity stars.  In a
recent work by Kewley \etal (2004) the metallicity dependence of
the \OII calibration was investigated  in great detail by examining the
\OII/\Ha ratio in empirical and theoretical approaches. They also find that
the calibration constant peaks at half solar metallicity, and their calibration constants
correspond fairly well to our values for Z=0.02, Z=0.008, and Z=0.004 with
values of 1.4, 2.0, and 1.3, respectively. Unfortunately, their calibrations cannot be used for nor extrapolated to our lowest and highest
metallicities, as their polynomial fit to the
\OII/\Ha vs. Z relation would then give unphysical negative values.     

Except for the lowest metallicity (1/50 solar), the metallicity effect on the calibration
of the \OII vs. SFR relation is small as compared to that of the \Ha vs. SFR relation. This
results from the counteracting effects of increasing metallicity and 
decreasing number of Lyc photons.  

We reemphasize the need to accurately correct an observed spectrum for dust  before using either \Ha or [OII] as a SFR indicator (cf. Kewley \etal 2004).  

\subsection{Impact of the choice of stellar evolutionary input physics}
The impact of the specific choice of stellar evolutionary input physics can be seen from a comparison between the present results based on recent Padova isochrones and those presented in Weilbacher \& Fritze - v. Alvensleben (2001) using Geneva stellar evolution models for the three metallicities in common, ${\rm Z_{\odot},~Z=0.008,~and~Z=0.004}$. Except for the stellar input physics, both approaches are identical, in particular, they both calculate the Lyman continuum photon fluxes on the basis of Schaerer \& de Koter's ${\rm N_{Lyc}(T_{eff})}$ calibrations. For solar metallicity, H$_{\alpha}$- and [OII]-fluxes are lower by 12 and 6\%, respectively, with Geneva than with Padova models, leading to emission line based SFR estimates higher by these percentages with Geneva than with Padova physics. Towards subsolar metallcities, the effect changes sign, and SFR estimates from H$_{\alpha}$ and [OII] are lower with Geneva than with Padova stellar evolutionary input physics by 9 and 4\%, respectively, for Z$=0.008$ and by as much as 25 and 30\%, respectively, for Z$=0.004$. 

\subsection{${\rm L_{UV}}$}

\begin{table}
\begin{center}
        \caption{Calibration constants for the UV luminosities vs. SFR relations. }
%        \begin{footnotesize}
%	\begin{sf}
        \begin{tabular}{r|rr|rr}\hline\hline
	 \multicolumn{1}{c}{${\rm M_{up}}$} & \multicolumn{2}{|c|}{100~\msun} & \multicolumn{2}{c}{120~\msun}\\
        \hline
	 Z& ${\rm C_{1500}}$& ${\rm C_{2800}}$ & ${\rm C_{1500}}$& ${\rm C_{2800}}$\\
	\multicolumn{1}{c}{} & \multicolumn{2}{|c}{$\left[{\rm \times10^{27}\frac{erg ~ s^{-1}~ Hz^{-1}}{M_\odot ~ yr^{-1}}}\right]$} & \multicolumn{2}{|c}{$\left[{\rm \times10^{27}\frac{erg ~ s^{-1} ~ Hz^{-1}}{M_\odot ~ yr^{-1}}}\right]$}\\
        \hline
          0.0004 & 14.6 & 13.7 & 14.8 & 13.9 \\
          0.004  & 12.8 & 11.8 & 13.1 & 12.0 \\
          0.008  & 11.9 & 10.8 & 12.2 & 11.0 \\
          0.02   & 10.6 & 10.0 & 11.0 & 10.3 \\
          0.05   &  9.3 &  8.9 &  9.8 &  9.2 \\
        \hline
       \end{tabular}
%       \end{sf}
%	\end{footnotesize}
       \label{calib2}
\end{center}
\end{table}

The UV luminosity  is also often used to derive SFRs. With the usable
wavelength range being 1250 - 2800 \AA{~}, a couple of calibrations are used
in the literature. We here use for our calibrations the wavelength ranges given by M+98 at
1500 \AA~ and 2800 \AA~ which are conventionally averaged over a rectangular bandpass of width
$\Delta\lambda/\lambda=20$\%.

\begin{displaymath}
{\rm L_{UV}~(erg~s^{-1}~Hz^{-1}) = C_{UV} \cdot SFR_{UV}~(M_\odot ~ yr^{-1})}.
\end{displaymath}
The calibration constants obtained with our models for galaxies with various
metallicities and different upper mass limits for the IMF are given in
Tab.~\ref{calib2}. In contrast to the emission line vs. SFR calibrations, these UV vs. SFR calibrations turn out to be almost insensitive to the upper mass limit of the IMF. Going from 100 to 120 ${\rm M_{\odot}}$ only increases the calibration constants by less than 2 and 4\% for ${\rm C_{1500}}$ and even less for ${\rm C_{2800}}$ at solar and low metallicities, respectively.  

Using Bruzual \& Charlot (1993, BC93) models with solar metallicity and a Salpeter IMF from 0.1 to 125 ${\rm M_{\odot}}$, M+98 found C$_{1500}=8.0 \times 10^{27}$ and  C$_{2800}=7.9 \times 10^{27}$ after transformation from their L$_{\nu}$ to our L$_{\lambda}$, i.e. 27 and 23\% smaller than our values, C$_{1500}=11.0 \times 10^{27}$ and  C$_{2800}=10.3 \times 10^{27}$, for an upper mass limit of 120 ${\rm M_{\odot}}$. The difference in the upper mass limit has negligible effect, the difference in the lower mass limit, however, accounts for 15\% of the difference. Another difference is that the
continuous emission of the gas is included in our models, but not in those used
by M+98. This causes a small increase in the UV luminosity for our models as
compared to the BC93 models that M+98 use. Neglecting the continuum emission in our models reduces our C$_{1500}$ and C$_{2800}$ by another 7\%. The remaining 5 and 1\% differences in C$_{1500}$ and C$_{2800}$ must be due to differences in the stellar evolutionary tracks and model atmospheres.
        
Like for \Ha vs. SFR, we find a strong metallicity dependence for the UV vs. SFR
calibrations as a result of the higher luminosities low metallicity stars. Additionally, their higher Lyc photon fluxes increase the continuous emission. 
Taking all effects into account, we find that the SFR of low metallicity galaxies is overestimated by  up to a 
factor 1.9 if using the M+98 calibration.

%______________________________________________________________ 
\section{Conclusions}\label{conclude} 

SFR determinations for galaxies on the basis of \Ha and \OII line luminosities,
and luminosities in the UV are common techniques used to estimate the SFR in
nearly all kinds of galaxies, from local ones to high redshifts. Even though
galaxies show a large scatter in metallicity, the standard calibrations are
derived from evolutionary synthesis using solar metallicity input physics.  We present here
a new set of calibrations for the most widely used SFR indicators \Ha, \OII,
and ${\rm L_{UV}}$ derived from evolutionary synthesis models and consistently
accounting for a large range in galaxy metallicities. Using recent Padova isochrones, the spectral library from Lejeune \etal (1997, 1998) and recent compilations of Lyc photon rates
as a function of stellar effective temperature (Schaerer \& de Koter 1997, Smith \etal 2002), we find good agreement with
standard calibrations in the literature for solar metallicity galaxy models. At
higher and lower metallicities, however, we find significant deviations.

Towards lower metallicities, in particular, we find strong deviations from the
SFR--\Ha and SFR--${\rm L_{UV}}$ relations and predict SFR of metal-poor galaxies
derived from standard solar metallicity calibrations to be overestimated by
factors up to 2 due to the hotter temperatures and the higher ionising fluxes and UV-luminosities of low-metallicity stars. 

In case of \OII, on the other hand, the lower oxygen abundance and the higher
output rate of Lyc photons act against each other and produce a maximum in
${\rm L_{[OII]}/SFR}$ around ${\rm Z=1/2~Z_\odot}$. Hence, the SFRs of galaxies around half
solar metallicity tend to be slightly underestimated while those of very low
metallicity galaxies are overestimated by the widely used standard
calibrations. Our results on [OII] agree well with those obtained from an independent approach by Kewley {\sl et al.} (2004).  

Our results depend on the choice of stellar evolution models as far as hot star temperatures and lifetimes are concerned. Differences between Padova and Geneva stellar evolutionary input physics are small at half-solar to solar metallicities ($\lta 10$\%), but increase to $\sim 30$\% towards Z$=0.004$. Stellar model atmosphere calculations have reached a comforting agreement in the last years as far as the UV- and the H-ionising fluxes are concerned, divergences only appear around He-ionisation.

Our results depend on the upper mass limit of the IMF with calibration constants for the \Ha and [OII] vs. SFR relations increasing by 13 and 20\% for low and solar metallicity, respectively, but only by less than 2 and 4\% for the UV vs. SFR constants in the sense that SFRs assuming too low an upper mass limit are overestimated by these percentages. Results also depend on the lower mass limit, since mass normalisation requires e.g. a higher number of low mass stars be compensated for by a smaller number of ionising UV-bright high mass stars. Going from 0.15 to ${\rm 0.1~M_{\odot}}$ lowers the Lyc and UV fluxes at a given SFR by $\sim 15$\%. 

Spectroscopy at all
redshifts picks out the most luminous and, hence, the most metal-rich galaxies. Hence, 
SFRs and SFR densities determined from \Ha and [OII] fluxes are expected to only be affected by metallicity effects at the highest redshifts. SFRs and SFR densities determined on the basis of restframe UV-luminosities of high redshift
galaxies in deep fields that also include the bulk of the lower luminosity and more metal-poor objects at each redshift, in contrast, are expected/predicted to be severely overestimated when using the standard calibrations.

We therefore stress the necessity to simultaneously determine the metallicity of
a galaxy and its SFR -- as can be done both from spectroscopy if enough emission
lines are seen and from multi-band imaging via the comparison between observed and model spectral energy distributions.

%______________________________________________________________ 
\begin{acknowledgements} 
This work was partly supported by the Deutsche Forschungsgemeinschaft (DFG) under 
grant Fr 916/10-1-2. We thank our referee, J. Gallagher, for his very helpful report.
\end{acknowledgements}


\begin{thebibliography}{} 
\bibitem[Anders \& Fritze-v.~Alvensleben(2003)]{2003A&A...401.1063A} 
Anders, P., \& Fritze-v.~Alvensleben, U., 2003, \aap ~401, 1063 
\bibitem[Bertelli et al.(1994)]{1994A&AS..106..275B} Bertelli, G., Bressan, 
A., Chiosi, C., Fagotto, F., \& Nasi, E., 1994, \aaps ~106, 275 
\bibitem[Bicker et al.(2004)]{2004A&A...413...37B} Bicker, J., 
Fritze-v.~Alvensleben, U., M{\" o}ller, C.~S., \& Fricke, K.~J., 2004, 
\aap ~413, 37 
\bibitem[]{} Brinchmann, J., Charlot, S., White, S. D. M., et al., 2004, \mnras ~351, 1151  
\bibitem[]{} Brinchmann, J. \& Ellis, R. S., 2000, \apj ~536, L77
\bibitem[Bruzual A.~\& Charlot(1993)]{1993ApJ...405..538B} Bruzual A., G., 
\& Charlot, S., 1993, \apj ~405, 538 
\bibitem[]{} Calzetti, D., 1997, \aj ~113, 162
\bibitem[Gallagher et al.(1989)]{1989AJ.....97..700G} Gallagher, J.~S., 
Hunter, D.~A., \& Bushouse, H., 1989, \aj ~97, 700 [G+89]
\bibitem[Inoue et al.(2001)]{2001ApJ...555..613I} Inoue, A.~K., Hirashita, 
H., \& Kamaya, H., 2001, \apj ~555, 613 
\bibitem[Izotov \& Thuan(1998)]{1998ApJ...500..188I} Izotov, Y.~I., \& 
Thuan, T.~X., 1998, \apj ~500, 188 
\bibitem[Izotov et al.(1994)]{1994ApJ...435..647I} Izotov, Y.~I., Thuan, 
T.~X., \& Lipovetsky, V.~A., 1994, \apj ~435, 647 
\bibitem[Izotov et al.(1997)]{1997ApJS..108....1I} Izotov, Y.~I., Thuan, 
T.~X., \& Lipovetsky, V.~A., 1997, \apjs ~108, 1 
\bibitem[Kennicutt(1992)]{1992ApJ...388..310K} Kennicutt, R.~C., 1992, 
\apj ~388, 310 
\bibitem[Kennicutt(1998)]{1998ARA&A..36..189K} Kennicutt, R.~C., 1998, 
\araa ~36, 189 [K98]
\bibitem[Kewley et al.(2004)]{2004AJ....127.2002K} Kewley, L.~J., Geller, 
M.~J., \& Jansen, R.~A., 2004, \aj ~127, 2002 
\bibitem[Kr\"uger et al.(1995)]{1995A&A...303...41K} Kr\"uger, H., 
Fritze-v.~Alvensleben, U., \& Loose, H.-H., 1995, \aap ~303, 41 
\bibitem[Lejeune et al.(1997)]{1997A&AS..125..229L} Lejeune, T., Cuisinier, 
F., \& Buser, R., 1997, \aaps ~125, 229 
\bibitem[Lejeune et al.(1998)]{1998A&AS..130...65L} Lejeune, T., Cuisinier, 
F., \& Buser, R., 1998, \aaps ~130, 65 
\bibitem[Madau et al.(1998)]{1998ApJ...498..106M} Madau, P., Pozzetti, L., 
\& Dickinson, M., 1998, \apj ~498, 106 [M+98]
\bibitem[]{} Mehlert, D., Noll, S., Appenzeller, I., 2003, Ap\&SS 284, 437
\bibitem[Mezger(1978)]{1978A&A....70..565M} Mezger, P.~O., 1978, \aap ~70, 
565 
\bibitem[Schaerer \& de Koter(1997)]{1997A&A...322..598S} Schaerer, D., \& 
de Koter, A., 1997, \aap ~322, 598 
\bibitem[Smith et al.(2002)]{2002MNRAS.337.1309S} Smith, L.~J., Norris, 
R.~P.~F., \& Crowther, P.~A., 2002, \mnras ~337, 1309 
\bibitem[]{} Stasinska, G., 1984, \aaps ~55, 15
\bibitem[Vacca et al.(1996)]{1996ApJ...460..914V} Vacca, W.~D., Garmany, 
C.~D., \& Shull, J.~M., 1996, \apj ~460, 914 
\bibitem[Weilbacher et al.(2000)]{2000A&A...358..819W} Weilbacher, P.~M., 
Duc, P.-A., Fritze-v.~Alvensleben, U., Martin, P., \& Fricke, K.~J., 2000, 
\aap ~358, 819 
\bibitem[Weilbacher \& Fritze-v.~Alvensleben(2001)]{2001A&A...373L...9W} 
Weilbacher, P.~M., \& Fritze-v.~Alvensleben, U., 2001, \aap ~373, L9 
\end{thebibliography}
\end{document}